\documentclass[aip,jmp,amsmath,amssymb,reprint,floatfix]{revtex4-1}
\usepackage{graphicx}
\usepackage{mathptmx}
\usepackage{graphicx}
\usepackage{amssymb}
\usepackage{color}
\usepackage[
   colorlinks,        
   linkcolor=blue,   
   filecolor=blue,   
   citecolor=blue   
]{hyperref}

\begin{document}


\title{Conducting and insulating LaAlO$_3$/SrTiO$_3$ interfaces: \\
A comparative surface photovoltage investigation}



\author{E. Beyreuther}
\email{elke.beyreuther@iapp.de}
\affiliation{Institut f\"ur Angewandte Photophysik, 
Technische Universit\"at Dresden, D-01062 Dresden, Germany}
\author{D. Paparo}
\affiliation{CNR-SPIN and Dipartimento di Fisica, Universit\`{a} di Napoli "Federico II", 
		Compl. Univ. di Monte S. Angelo, via Cintia, I-80126 Napoli, Italy }
\author{A. Thiessen}
\author{S. Grafstr\"om}
\author{L. M. Eng}
\affiliation{Institut f\"ur Angewandte Photophysik, 
Technische Universit\"at Dresden, D-01062 Dresden, Germany}


\date{\today}

\begin{abstract}
Surface photovoltage (SPV) spectroscopy, which is a versatile method to analyze the 
energetic distribution of electronic defect states at surfaces and interfaces of wide-bandgap 
semiconductor (hetero-)structures, is applied to comparatively investigate 
heterostructures made of 5-unit-cell-thick LaAlO$_3$ films grown either on TiO$_2$- 
or on SrO-terminated SrTiO$_3$. As shown in a number of experimental and 
theoretical investigations in the past, 
these two interfaces exhibit dramatically different properties with the first being 
conducting and the second insulating. 

Our present SPV investigation reveals clearly distinguishable 
interface defect state distributions for both configurations when interpreted 
within the framework of a classical semiconductor 
band scheme.  Furthermore, bare SrTiO$_3$ crystals with TiO$_2$ or mixed SrO/TiO$_2$ 
terminations show similar SPV spectra and transients as do LaAlO$_3$-covered 
samples with the 
respective termination of the SrTiO$_3$ substrate. 
This is in accordance with a number of recent works 
that stress the decisive role of SrTiO$_3$ and the minor role of LaAlO$_3$ 
with respect to the electronic interface properties.
\end{abstract}

\pacs{68.08.-p, 68.47.Gh, 71.20.Ps, 71.30+h, 72.20.-i, 72.40.+w, 
	73.20.-r, 73.40.-c, 73.50.Pz, 73.61.Ng, 77.84.Bw}
\keywords{strontium titanate, SrTiO$_3$, lanthanum aluminate, LaAlO$_3$, perovskite, interface, 
surface photovoltage, SPV, interface state, surface photovoltage spectroscopy, Kelvin probe}

\maketitle

\section{Introduction}
\label{Introduction}

Perovskite oxide heterostructures in general and their interface physics in particular have 
attracted a lot of scientific interest in the past and are one key issue of the highly 
topical field of \emph{oxide electronics}, which is dedicated to finding novel material systems and 
device concepts for micro- and nanoelectronics beyond the conventional semiconductor technology. 
In this context the LaAlO$_3$/SrTiO$_3$ (LAO/STO)
system has been the subject of intense research activities, especially since the publication 
of Ohtomo and Hwang in 2004 \cite{oht04}, in which the existence of a conducting 
2-dimensional electron gas (2-DEG) -- in the meantime more frequently 
characterized as electron liquid (2-DEL) -- at the 
interface between TiO$_2$-terminated SrTiO$_3$ and LaAlO$_3$ thin films was 
described. The conducting interface occurs only in samples with LAO layers 
thicker than 4 unit cells and never for samples with SrO 
termination of the STO substrate. Until now, this work has stimulated a vast 
number of theoretical and experimental works on this prototypical LAO/STO system -- several reviews 
\cite{mau07,pau08,cha10,zub11} have been published, too -- to clarify 
the origin and the preconditions of the appearance of either a conducting or an insulating 
interface. Furthermore, it was not only the unexpected conductivity 
of the interface that attracted such a big interest, but also a number of subsequent works showing 
intriguing properties of the 2-DEL including 
superconductivity \cite{rey07} and its manipulation by an external electrical 
field \cite{cav08}, large negative magnetoresistance \cite{bri07}, the tunability of the 
conduction state on the nanoscale by an atomic force tip \cite{cen08,cen09}, photoresponse 
on the nanoscale \cite{irv10}, the coexistence of ferromagnetism 
and superconductivity \cite{dik11,li11}, electrostriction \cite{can11}, 
or charge and orbital order \cite{rub11,ris11}. They all make the system a model system for 
studying very fundamental solid state phenomena confined to an interface.

The origin of the charge carriers has been a much debated question, since 
different intrinsic and extrinsic doping mechanisms can be at play in this 
oxide heterostructure. Many recent results point to the
so-called polar catastrophe scenario as the mechanism driving the formation 
of the conducting state \cite{can11,thi06,pau11,rei12,liu13}. According to 
this picture, the polar discontinuity, which occurs at the interface
between the charge-neutral planes of SrO and TiO$_2$ and the charged LaO 
(+e, where e is the electron charge) and AlO$_2$ (--e) sheets, causes 
an electrostatic breakdown once the LaAlO$_3$ layer has reached a critical thickness. 
However, over the years a lot of experimental data 
have been gathered suggesting that this picture might be oversimplified and 
that the real structure of the interface 
including oxygen vacancies, lattice distortions, and cation intermixing also determine the 
conduction properties \cite{cha11}. 

In particular, we mention a striking 
phenomenon connected to the specific termination of the substrate, namely 
the prevention of the 2DEG formation when the STO substrate is terminated 
with a SrO plane. In this case, not considering
possible interfacial reconstructions, the SrO-terminated system is structurally 
identical to the TiO$_2$-terminated one, except for the two half-unit cells 
adjacent to the interface on the LAO and STO sides. 
However, this slight difference is sufficient to have dramatic 
consequences on the electronic-
transport and optical-spectro\-scopic properties of the interface \cite{rub13}.
Since the electronic transport within heterostructures is decisively 
determined by the distribution and properties of interface defect states, 
a comparative characterization of these LAO/STO interfaces with the 
two different STO terminations in terms of defect states might clarify the 
origins of the extremely unequal conductivities. Before we describe our approach 
to characterize the electronic interface structure,
which is based on the analysis of the surface photovoltage of the two 
LAO/\-STO systems, we shortly discuss the existing literature dedicated to spectroscopic 
defect state investigations of LAO/\-STO.

\subsection{Spectroscopy of electronic in-gap states at LaAlO$_3$/SrTiO$_3$ interfaces -- 
State of the art}

Despite the high total number of publications on the LAO/STO topic, there has been only 
a manageable amount of work 
focusing on the direct analysis of the electronic (defect) structure of this interface: 

\subparagraph{Scanning tunneling microscopy/spectroscopy}
The STS investigation by Breitschaft et al. \cite{bre10} aims at identifying the nature of the conducting 
interface electrons with respect to their density of states. A band scheme 
is derived and it is stated that the 
situation of the interface electrons is very different than in the case of 2-dimensional electron 
gases in III-V semiconductor heterostructures. In the latter, the band bending makes potential wells, 
which trap and localize carriers at the interface, but for LAO/STO a classical band 
scheme seems to fail: A more ionic picture, in which the electrons exist in narrower potential 
wells made up by the Coulomb potential of the Ti ions in the TiO$_6$ octahedra of the perovskite structure, 
and in which band bending does not play the main role, is suggested. 

Another STM/STS study \cite{ris12} compares 4~u.c.-thick (conducting) to 2~u.c.-thick (insulating) 
LAO films on TiO$_2$-terminated STO -- however, the samples were illluminated to make them 
conductive enough for STM by photodoping -- and finds intrinsic in-gap states in both specimens as well 
as nanoscale electronic inhomogeneities induced by spatially localized electrons. This work 
supports the view that we are facing a system on the verge between an ionic and a band picture here.

\subparagraph{Soft-x-ray spectroscopy}
Soft-x-ray spectroscopy was used by Drera et al. \cite{dre11} to compare 
a 3-u.c. and a 5-u.c. LAO film, both on TiO$_2$-terminated STO. Indeed,
in-gap states at the LAO/STO interface were found. They were shown to have Ti 3d character and 
their concentration was higher in the case of the insulating interface.

In the extended comparative x-ray absorption investigation by Koitzsch et al.~\cite{koi11} 
a strong dependence of the in-gap state configuration on the oxygen pressure and the 
deposition methods(PLD or MBE) is proposed. In-gap states were only found for the 
PLD-grown samples but not for the MBE-grown ones. In the former, one 
O-2p-related and two Ti-3p-related states were detected and discussed.

However, recent results of Cancellieri et al. \cite{can13}, comparing the valence 
band spectra of 2.5-u.c. and 4-u.c. LAO films, showed, in contrast to Drera et al., 
no evidence for in-gap states but clear signatures for states at the Fermi edge 
exclusively for the 4-u.c. sample with the conducting interface.
Hitherto, an unambiguos picture from photoemission spectroscopic investigations 
is still missing.

\subparagraph{Optical absorption}

The work of Wang et al. \cite{wan11} 
contains optical absorption data of a 26-u.c. LAO film on TiO$_2$-ter\-mi\-nated STO. The 
authors find three distinct features in the optical absorption spectrum, which are all associated 
with defect levels of the STO. The features were found \emph{only} in samples grown under 
sufficiently low oxygen pressure.

\subparagraph{Conclusions}

Up to now, several independent studies, employing different experimental techniques, 
have been focused on detecting and/or mapping the distribution of in-gap interface states 
within LAO/STO heterostructures have been performed. So far, there is no overall 
coherent picture due to the incomplete comparability of the methods and the samples.

In the present paper, we present an alternative approach: the mapping of electronic interface 
states
by analyzing the surface photovoltage of the system. In contrast to many previous studies, 
we compare the interface state distribution as a function of the STO termination and not 
of the LAO thickness. 
Since SPV methods are not as frequently used 
as STM, XAS, or optical absorption, some methodological details are given in the 
following. 

\subsection{Surface photovoltage phenomena for interface characterization}

SPV phenomena were already known during the pioneering period of semiconductor physics, i.e., 
back in the 1940s and 1950s (see e.g. \onlinecite{joh58}). 
SPV in classical semiconductors such as Si 
and Ge is well understood and industrially employed for the contamination analysis of wafers 
\cite{lag92,schro82}. For wide-bandgap materials, for which purely electrical characterization 
methods fail due to low carrier concentrations -- and this applies also to many perovskites --, 
SPV has been found to be a valuable tool for surface gap state spectroscopy since the 1970s 
\cite{gat73,lag94}.

\subparagraph{Information content}

An SPV analysis is capable of providing information on the energetic defect distribution at 
a given -- optically accessible -- interface and on a variety of quantities, e.g., the carrier
diffusion length, surface band bending, surface charge, and surface dipole, as well 
as surface and bulk recombination rates, distribution and properties (time constants 
and optical cross sections) of surface states, distinction between surface and bulk 
states, conduction type, or construction of band diagrams. 
The topic was extensively reviewed by Kronik and Shapira \cite{kro99,kro01}.

\subparagraph{Physical background}

The physical principle behind the SPV is commonly explained employing a classical 
semiconductor band scheme of the surface or interface of interest. In brief, the change 
of the surface band bending of a semiconductor 
surface or interface under illumination, as induced by charge carrier redistribution, 
is denoted as surface photovoltage. Depending on whether 
the SPV is generated 
by excitation with photon energies below or above the bandgap the terms 
\emph{sub-} and \emph{super-bandgap SPV} are used. The first case allows us to 
induce a direct charge transfer between energetically localized in-gap states 
and one of the bands. In practice, the SPV is recorded as a function of excitation 
energy and the points of slope change in such a \emph{SPV spectrum} can be 
associated with the energetic positions of the respective in-gap states.

\subparagraph{SPV at oxidic interfaces}

While -- especially sub-bandgap -- SPV investigations have been frequently 
performed to analyze surfaces of wide-gap 
semiconductors \cite{gat73} belonging to the II-VI (such as CdS, CdSe) 
and the III-V (GaAs, GaN, InP)
group, as well as a number of inorganic/inorganic and organic/inorga\-nic 
interfaces, solar cell 
structures, light-emitting materials, and quantum wells and dots \cite{kro99}, 
works focusing exclusively on oxidic materials are rare 
so far: TiO$_2$ \cite{duz01b,li02}, ITO \cite{li02}, SrTiO$_3$ \cite{mav78,bey13}, and 
man\-ganite/SrTiO$_3$ as well as gold/ \linebreak SrTiO$_3$ 
interfaces \cite{bey11} were examined. The latter two can be seen as preliminary 
investigations of the present work.

\subsection{Motivation and outline of the present work}

So far, extended comparative
investigations of the LAO/STO system's \emph{electronic surface and interface defect distributions} 
are widely lacking. In the current work an \emph{optical} approach utilizing 
surface photovoltage phenomena, is employed to investigate two 5-u.c.-thick LAO films on 
either TiO$_2$- or SrO-terminated STO as well as the respective bare STO single crystals. 
After a short description of the samples and the experiments, wavelength- 
and time-dependent SPV measurements performed on all four samples are presented. 
The value of the SPV method for investigating wide-bandgap heterostructures such as LAO/STO is discussed and 
possible directions of further useful experiments are suggested.

\section{Experimental}
\label{Experimental}

\subsection{Samples}

The SPV investigations were performed with two different \emph{LAO/STO heterostructure samples}:
\begin{itemize}
\item 5 unit cells of LaAlO$_3$ grown on TiO$_2$-terminated SrTiO$_3$ ({\bf LAO/TiO$_2$:STO}), 
	exhibiting a \emph{conducting} interface with a sheet conductance of 
	$\sigma_S = 0.05$~mS at 300~K,
\item 5 unit cells of LaAlO$_3$ grown on SrO-terminated SrTiO$_3$ ({\bf LAO/SrO:STO}),
	exhibiting an \emph{insulating} interface ($\sigma_S$ below the detection limit 
	of 10$^{-9}$~S);
\end{itemize} 
the "corresponding" two \emph{bare SrTiO$_3$ (001) single crystals}:
\begin{itemize}
\item SrTiO$_3$ with TiO$_2$ termination ({\bf TiO$_2$:STO}),
\item SrTiO$_3$ with mixed SrO/TiO$_2$ termination ({\bf mixed:STO});
\end{itemize}
as well as (for selected further comparative measurements)
\begin{itemize}
\item a bare LaAlO$_3$ (100) single crystal\footnote{Of course, a LAO bulk crystal cannot act as 
a real reference for a 5-u.c.-thick thin film, since its physical properties might be decisively different. On 
the other hand it is not possible to check the SPV behavior of a free-standing film. Thus the usage 
of the bulk crystal is a compromise, also because there is no SPV data on LAO available in 
literature so far.} ({\bf LAO}).
\end{itemize}

The reason for contrasting the LAO/SrO:STO sample with the mixed:STO sample instead of 
a SrTiO$_3$ single crystal with pure SrO termination is that the latter is 
not stable in air.

To obtain a single TiO$_2$termination, as-received (i.e., mixed-terminated) 
SrTiO$_3$ substrates were prepared according to the procedure reported in 
\onlinecite{kos98}: The purchased substrates were ultrasonically
cleaned in demineralized water for 10~min and dipped for
about 30~s into a commercially available buffered fluoridic acid
solution. To facilitate surface recrystallization and remove any
residual contamination, a final annealing step was performed
at 950$\symbol{23}$C for several hours in flowing oxygen. 
SrO-terminated substrates were obtained
by depositing a SrO layer on a TiO$_2$-terminated STO crystal 
before growing in-situ the LAO
film \cite{rub13}. 
The SrO monolayer is grown by means of the same technique used for 
depositing LAO films, as explained below. 

The LAO films of five unit cells thickness were prepared by pulsed
laser deposition (PLD) on (001)-oriented STO substrates with
uniform TiO$_2$ or SrO terminations. The
thickness of the layers was monitored during the deposition
using reflection high-energy electron diffraction (RHEED)
oscillations. This allowed us to precisely control the layer-by-layer growth. 
The samples were grown at 800$\symbol{23}$C in an oxygen atmosphere of
1$\times$10$^{-4}$~mbar; annealing
in 0.2~bar of oxygen at 550$\symbol{23}$C for one hour preceded the
cool-down to room temperature at the same pressure \cite{can11}.

The samples were 5$\times$5$\times$0.5~mm$^3$ in size with unpolished rear sides, which were 
covered with silver paint to mount and contact the samples within the Kelvin probe device that was 
employed to measure the SPV.

\subsection{Surface photovoltage measurements}

\subparagraph{Setup}

In order to measure the SPV as a function of the illumination wavelength 
as well as to record SPV transients after switching the light on and off, we combined 
a home-built  Kelvin probe with an illumination setup. 
The latter uses a monochromatized Xe arc lamp (providing a wavelength range between 
220 and 800~nm) as the light source. 
The Kelvin probe measures the contact potential difference (CPD) between the sample 
surface and a reference gold electrode, which oscillates perpendicularly to the sample surface. 
The SPV is the difference of the CPD under illumination and in the dark.
It is thus 
a differential quantity independent of the absolute work function of the electrode material. 
For a more detailed description of the Kelvin probe and the 
illumination setup, refer to \onlinecite{bey11}.
 
During \emph{wavelength-dependent} SPV recording, the photon flux was kept constant by adjusting 
the light intensity by means of a motorized neutral density filter. For the 
measurements of \emph{SPV transients} the light beam was blocked or released via a PC-controlled shutter. 
All measurements were carried out under ambient conditions, but with stabilized 
temperature.

\subparagraph{SPV spectroscopy}

The SPV spectra were collected in four (overlapping) spectral ranges, motivated 
by practical reasons (changes of edge filters and monochromator gratings): 800--550~nm, 
600--410~nm, 450--320~nm, 340--220~nm. The overall spectral range is in 
accordance with the band gaps: 800~nm is around half of the STO gap 
(387~nm, 3.2~eV), 220~nm is nearly the LAO 
band gap (221~nm, 5.6~eV). The photon fluxes were 
chosen as follows: 5$\times$10$^{10}$~s$^{-1}$ for the 450--220~nm 
range\footnote{Using two different photon fluxes was motivated as follows: 
The photon flux is limited by the output power 
of the Xe arc lamp. The latter drops dramatically in the ultraviolet range. Thus, the maximum 
achievable flux at 220~nm determines the flux for the whole spectrum. On the other hand, at the 
other end of the spectral range, on the long-wavelength side -- far away from the STO and 
LAO bandgaps -- the SPV response is quite low. Thus we decided to use a slightly higher photon 
flux for the two long-wavelength spectra.} and 
4$\times$10$^{11}$~s$^{-1}$ for the 800--410~nm range. The illuminated area measured
around 2$\times$10$^{-5}$~m$^2$ (corresponding to the area of the semitransparent reference 
electrode of the Kelvin probe). 

Prior to the recording of any SPV spectrum the sample was kept in dark for at least 20~hours. 
The desired photon flux 
density at the starting wavelength (e.g. 800~nm) of the respective partial spectrum was adjusted. 
Still in the dark, the dark value of the contact potential difference $CPD_{dark}$ was recorded. Then, the beam path towards the sample was unblocked and the 
CPD was monitored until it had stabilized. Since the initial transient behavior 
of the SPV strongly varies for the different starting wavelengths of the four partial 
spectra, this waiting time differs typically between one and ten minutes. Subsequently
the actual SPV-vs.-wavelength data acquisition started. The wavelength was swept in steps of 
1~nm with a waiting time of 30~seconds before the respective CPD value was recorded. 
The SPV values were calculated as $SPV=CPD_{dark}-CPD_{illuminated}$.

\subparagraph{SPV transients}

\begin{figure}
   \centering
   \includegraphics[width=0.48\textwidth]{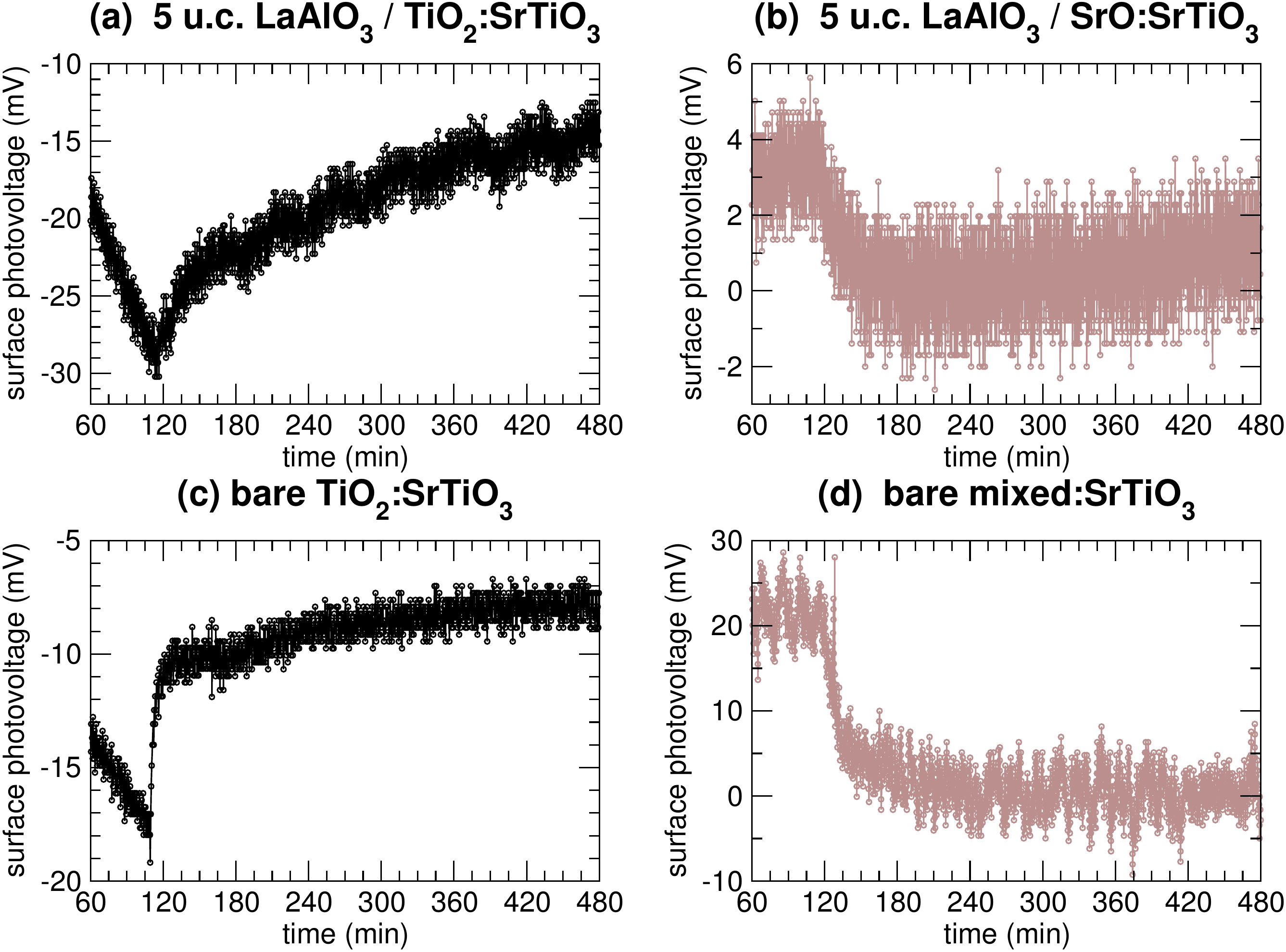} 
   \caption{SPV transients after illumination with 600-nm light for 
		$\approx$100~minutes. After the 
		exposure time, the light was blocked and the SPV relaxation was recorded for several 
		hours. The first 2~hours of the \emph{light-off} transients are shown for 
		(a), (b) the heterostructures and (c), (d) the STO reference samples.}
   \label{LAO_STO_comp_transients}
\end{figure}

At selected wavelengths, namely at the four starting wavelengths of the partial spectra 
(800~nm, 600~nm, 450~nm, 340~nm), the SPV was also monitored as a function of time 
at the same photon fluxes as used during the spectral measurements. In detail, the transients 
were acquired as follows: As in all cases, the initial dark value was recorded; then the light beam 
was unblocked and the SPV was recorded for the same time span as the respective spectrum 
had taken (e.g., around 100~minutes at 600~nm, see figure~\ref{LAO_STO_comp_transients}). 
This part of the transient is called \emph{light-on transient}. Subsequently, the beam was blocked 
and the SPV recording continued for several hours in order to monitor the 
relaxation of the SPV -- denoted \emph{light-off transient}. There were a number of reasons for measuring transients. 
First, it had to be tested whether some very-long-term relaxation features were present and caused 
artefacts in the SPV spectra. In fact, this effect occurred only once, as discussed 
in the supplement\footnote{See supplemental material at 
\url{http://www.iapp.de/~elke/preprints/Beyreuther_et_al_2013_LAO_STO_SPV_supplement.pdf}
for additional diagramms of 
the SPV data.} 
(figure~S1). Second, 
the transients represent essential complementary data for comparing 
the electronic properties of the 
samples, since for SrTiO$_3$-based heterostructures the 
a full explanation and interpretation of the purely spectral data alone 
is commonly not unambigously possible \cite{bey11}. 

\section{Results and discussion}
\label{Results}

\begin{figure}
   \centering
   \includegraphics[width=0.48\textwidth]{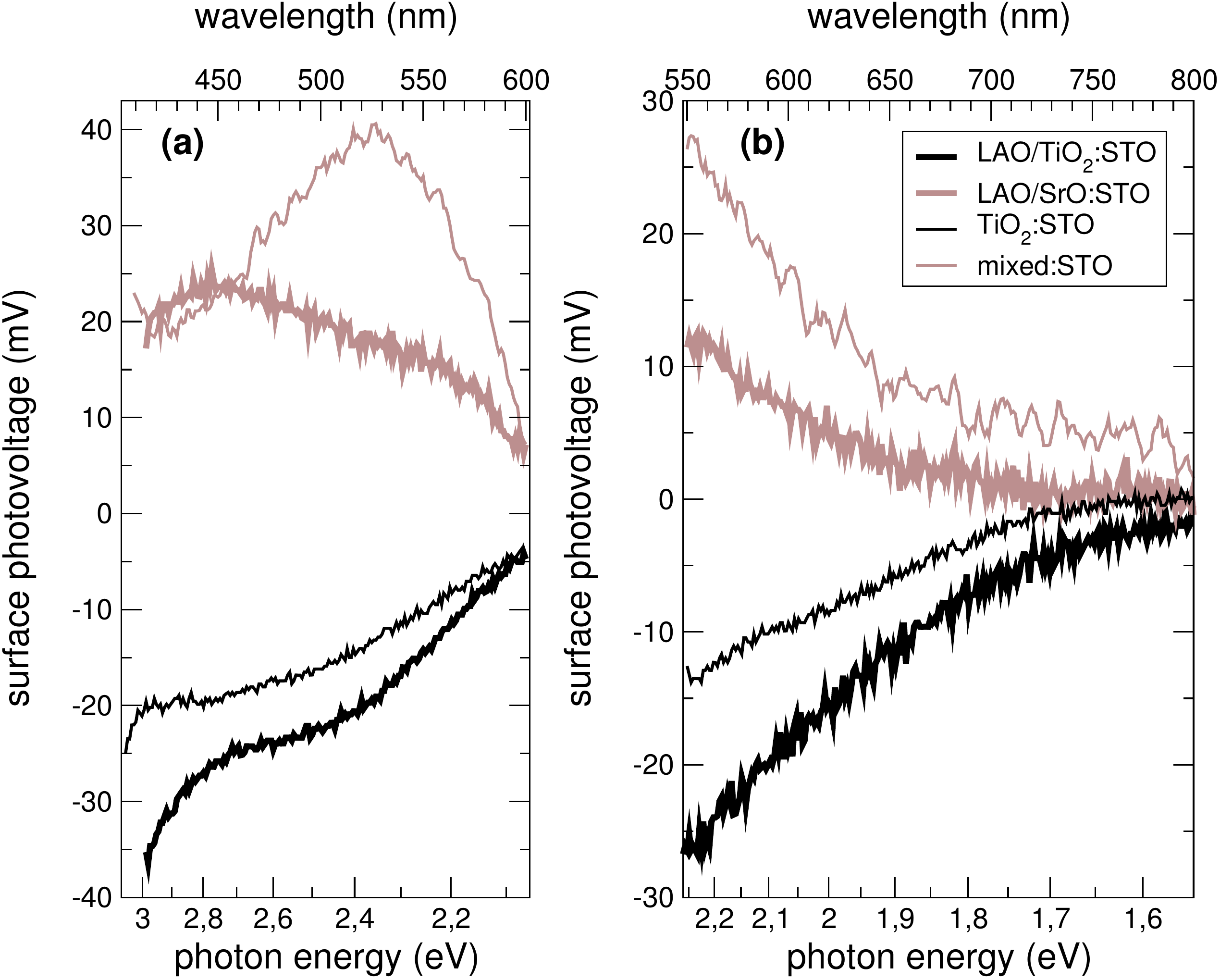} 
   \caption{Overview of the \emph{sub-bandgap} surface photovoltage spectra of 
   both the LAO/STO heterostructures (thicker lines) and the bare STO 
   reference samples (thinner lines), 
   recorded within two overlapping spectral ranges: (a) 600--400~nm; 
   (b) 800--550~nm.}
   \label{LAO_STO_comp_spectra_sub}
\end{figure}

\begin{figure}
   \centering
   \includegraphics[width=0.48\textwidth]{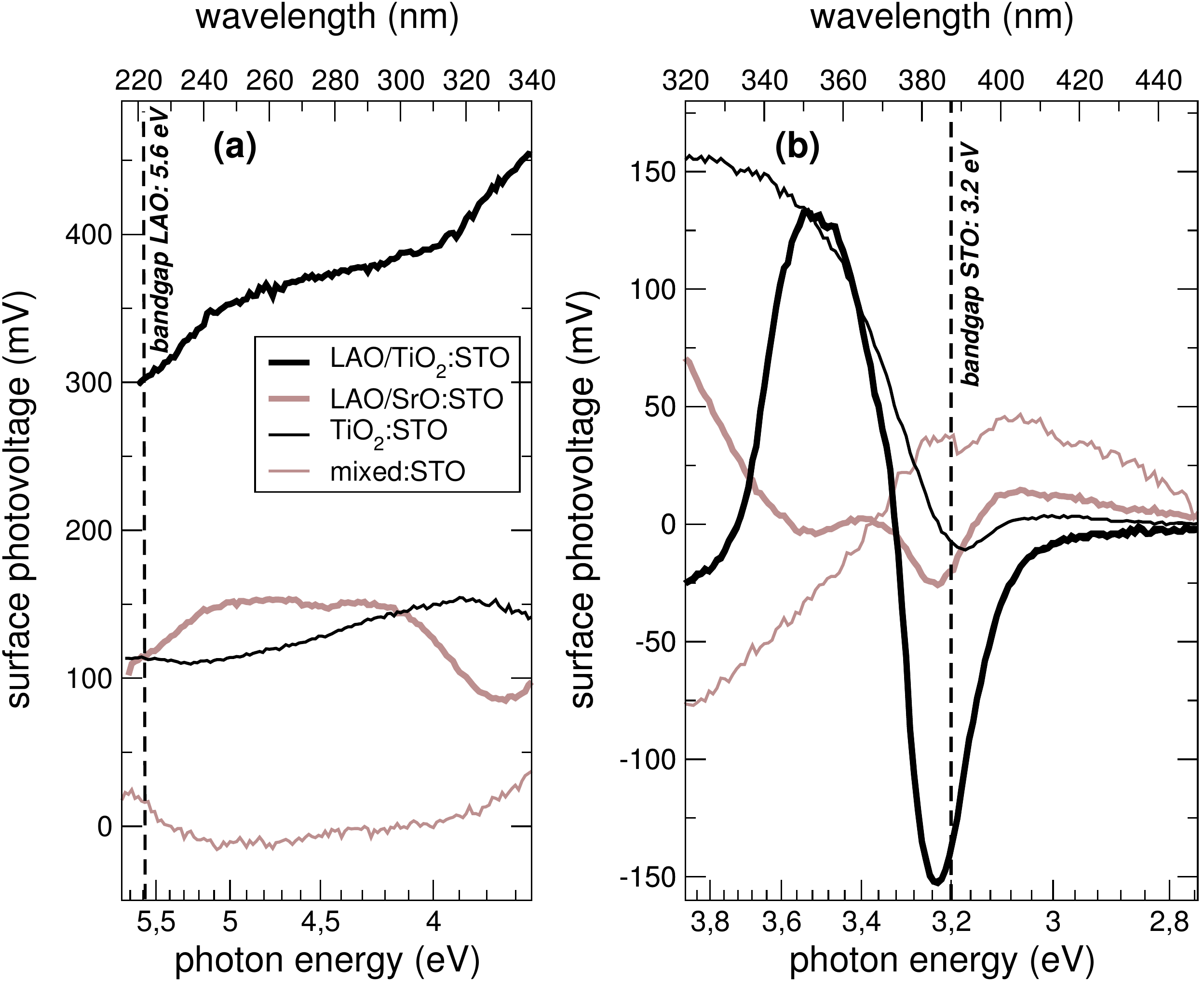} 
   \caption{Overview of the surface photovoltage spectra of 
   both the LAO/STO heterostructures (thicker lines) and the 
   bare STO reference samples (thinner lines), 
   taken across STO's \emph{super-bandgap} region above 3.2 eV and 
   up to the LAO bandgap at 5.6 eV. For technical reasons, the data 
   was recorded in two overlapping spectral ranges: (a) 340--218~nm, 
   (b) 450--320~nm.}
   \label{LAO_STO_comp_spectra_super}
\end{figure}

\begin{figure}
   \centering
   \includegraphics[width=0.48\textwidth]{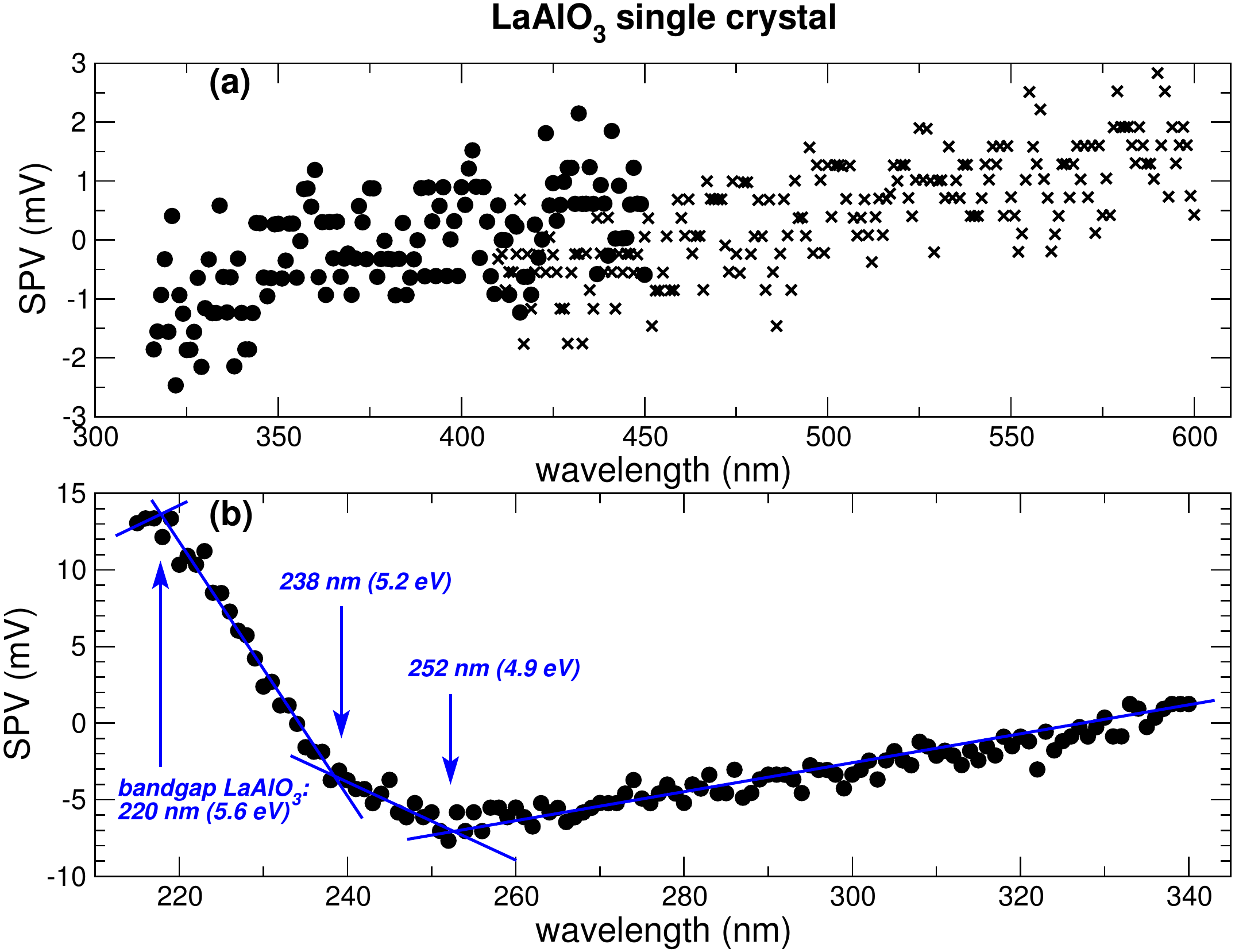} 
   \caption{Sub-bandgap SPV spectra of a bare LAO (100) single crystal: (a) Besides a 
	slight drift, there is no photoresponse in the range between 600 and 320~nm. The 
	different data point symbols indicate that the spectrum was taken within two overlapping 
	parts, as explained in the text. The noise band is around 3 mV. (b) In the range between 
	340 and 220~nm higher SPV values as well as three points of slope change can be 
	stated.}
   \label{LAO_spectra}
\end{figure}

\subsection{General features}

Let us start with a discussion of the general shape of the
\emph{sub-bandgap}\footnote{In the following we use the 
term \emph{sub-bandgap} with respect to the band gap of SrTiO$_3$ (for LaAlO$_3$ the whole 
spectral range covered by our light source is sub-bandgap).}
surface photovoltage spectra, as shown in 
figure \ref{LAO_STO_comp_spectra_sub}. There, the thick lines represent the two 
LAO/STO heterostructure samples and the thinner lines stand for
the bare STO reference samples, respectively. 

Considering the LAO/TiO$_2$:STO 
and the LAO/SrO:STO spectra first, we notice that both spectra are clearly different as 
they exhibit (i) higher absolute values of the SPV for the sample with the 
conducting interface, which might be directly linked to the higher interface 
conductivity, as well as (ii) different signs of the SPV. A detailed microscopic 
clarification of the origin of the 
different signs would be too speculative, since this would 
require a concrete idea of the band lineup at the two different LAO/STO interfaces. 
To date there exists no consensus regarding the
proposed band alignments, the prediction and measurement of the built-in fields, 
band offsets, band bending, or band discontinuities 
\cite{can11,rei12,cha11,bre10,qia11,yos08,gud13}.

More illuminative is the comparative view on the spectra of the bare STO samples, which 
serve as a reference. Obviously, the TiO$_2$-terminated STO sample behaves very similar 
to the LAO/TiO$_2$:STO heterostructure and the STO sample with 
mixed SrO/TiO$_2$ termination behaves more like the LAO/SrO:STO heterostructure, 
at least with respect to the sign of the STV. Regarding the fact that it is not a pure SrO termination, this 
is quite astonishing. Thus, from this rough analysis, it is 
mainly the STO's termination, which determines the sub-bandgap SPV response and the 
related defect state distribution and dynamics.

Finding such general shape similarities between the \emph{su\-per-bandgap} spectra 
of the four samples, as depicted in figure \ref{LAO_STO_comp_spectra_super}, is 
not equally possible, since each of the samples shows quite individual spectral features, 
apart from the STO band gap being visible in all of the spectra. Part of the super-bandgap 
features might directly originate from higher optical transitions 
(cf.~\onlinecite{yac73}), others may stem from the convolution of the SPV spectrum 
with multi-level 
relaxation processes, which are typical for SrTiO$_3$ under super-bandgap excitation 
\cite{bey13,bey11}, or persistent photoconductivity effects \cite{teb12}.

The clear shape similarities that were visible in the sub-bandgap SPV spectra 
are further evidenced by the light-off SPV transients of the four samples, see 
figure~\ref{LAO_STO_comp_transients}, with the same sample pairs showing 
similar shapes.

Further, though weaker, evidence for the major role of the STO physics and the 
minor role of the LAO for the electronic defect state distribution and dynamics, can be 
derived from the SPV spectra of the bare LAO single crystal, as depicted in 
figure~\ref{LAO_spectra}. Between 600 and 300~nm there is almost no SPV response, and from 
300 to 220~nm the SPV is much lower than in the STO-based samples.

\begin{figure*}
   \centering
   \includegraphics[width=0.75\textwidth]{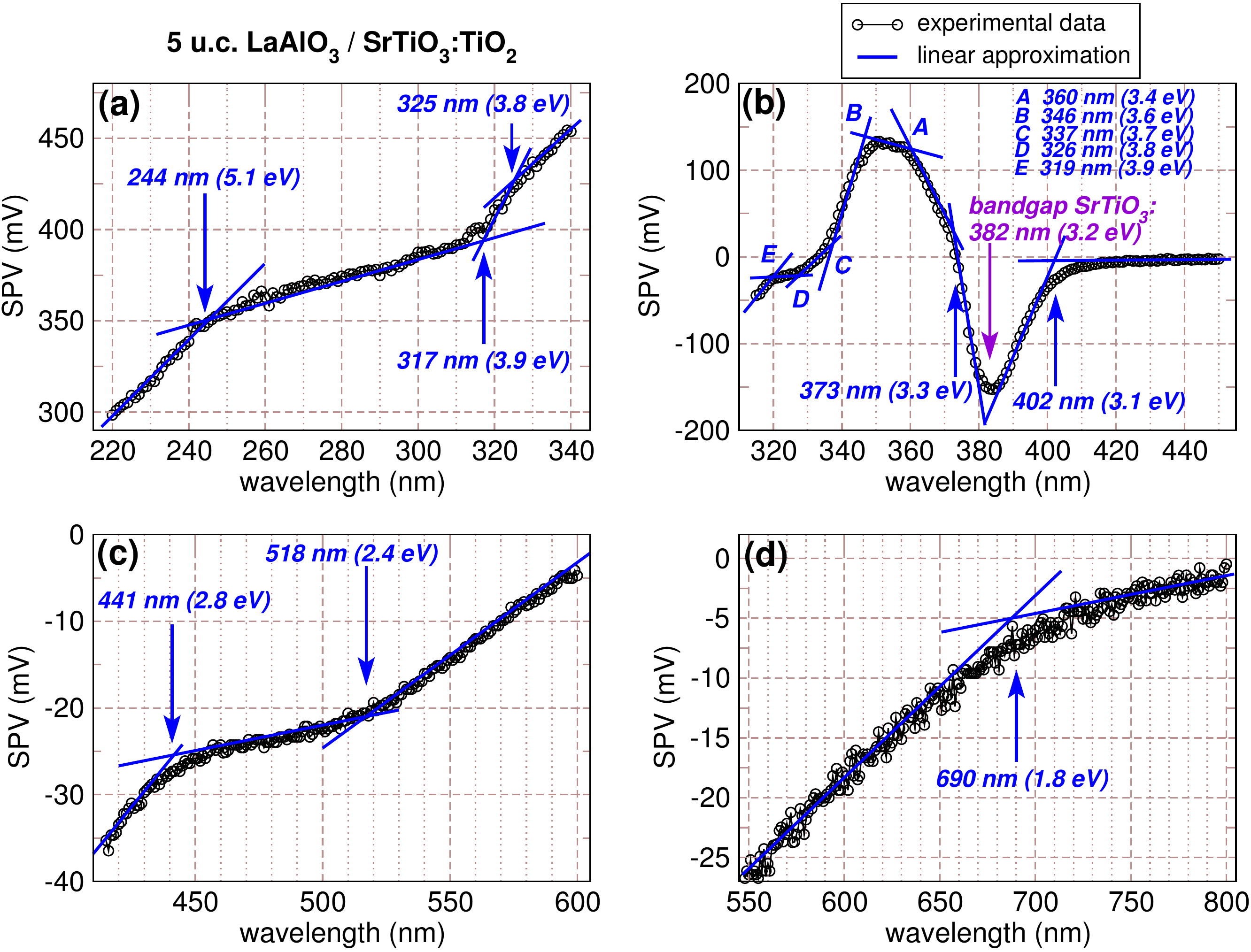} 
   \caption{Surface photovoltage spectra of the LAO/TiO$_2$:STO sample (conducting interface), 
	taken between 800 and 220~nm within four overlapping wavelength ranges: (a)-(d). The 
	points of slope changes, which indicate the beginning of filling or emptying an energetically 
	localized in-gap state, are marked by arrows. The corresponding data for the other three 
	samples are shown in the supplement as figures S3--S5.}
   \label{fig_LAO_TiO2_STO_slope_changes}
\end{figure*}

\subsection{Spectral features}

\begin{figure*}
   \centering
   \includegraphics[width=0.75\textwidth]{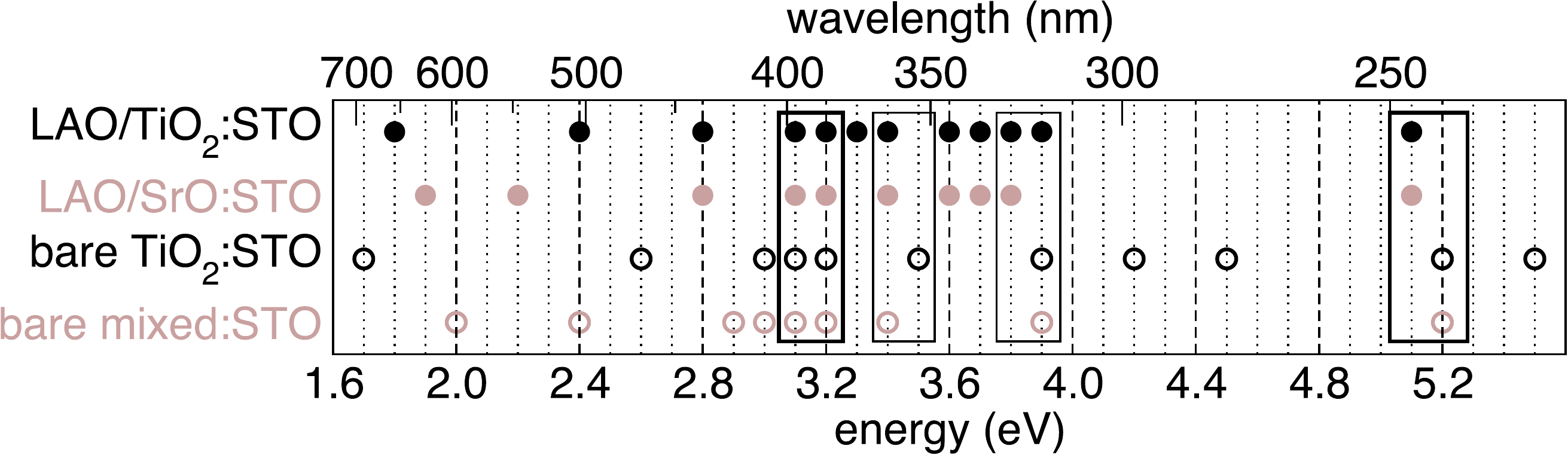} 
   \caption{Overview of all points of slope changes, i.e., all in-gap states 
		of the LAO/STO and bare STO samples with their energetic positions.}
   \label{LAO_STO_comp_energy_positions}
\end{figure*}

We continue with a more detailed examination of the several spectral features. As 
described in the introductory part, points of slope changes in sub-bandgap SPV spectra 
are commonly associated with  the energetic positions of electronic defect states 
located within the band gap. Figure~\ref{fig_LAO_TiO2_STO_slope_changes} shows 
the full analysis of the SPV spectra of the LAO/TiO$_2$:STO heterostructure. Points 
of slope changes are derived from the intersection points of the linear approximations 
of the parts with constant slope. The corresponding wavelengths and photon 
energies are given at the respective arrows. The SPV spectra of the LAO/SrO:STO and 
the two bare STO crystals were analyzed in the same manner. The corresponding 
figures have been shifted to the supplement (figures S3--S5). For all four samples, 
the extracted energetic positions of the points of slope change are summarized 
in figure~\ref{LAO_STO_comp_energy_positions}. At least for the sub-bandgap 
range of STO, i.e. up to 3.2~eV, this graphics represents a \emph{map} of the in-gap 
states, which will be discussed in the following:

Within the sub-bandgap range no features common to all four samples were observed. 
In contrast to the overall shape (see figure~\ref{LAO_STO_comp_spectra_sub} again), 
all four in-gap state distributions 
are different. Only the feature at 3.1~eV shortly below the STO bandgap, 
which can be attributed to shallow states and/or a pronounced Franz-Keldysh effect as often 
observed in SPV spectra, and the feature at 3.2~eV due to band-to-band excitation of carriers 
are, as expected from our former SPV results on manganite/STO heterostructures \cite{bey11} 
and STO single crystals \cite{bey13}, present in all samples.

The feature at 2.8~eV (approx.~440~nm) appears only in the two LAO-covered 
samples and is probably the same level as observed by Wang et al.~\cite{wan11}. 
There, the level was explained as "self-trapped exciton state/impurity level" 
according to former works on STO single crystals \cite{bae66,lee75,kan05}. 
Here, this level might be induced by the LAO deposition process, 
which possibly modified the STO defect configuration, since none of the bare STO samples 
shows this feature. In this context it would be highly interesting to study 
the evolution of this level 
in gradually more defective samples or in a sample 
series with growing number of LAO unit-cell layers.

The features at 1.8~eV (LAO/SrO:STO) and 1.9~eV (LAO/ TiO$_2$:STO) probably have 
the same origin -- a deep oxygen vacancy defect level -- and were also observed in the 
optical absorption measurements by Wang et al.~\cite{wan11}, while the third 
pronounced in-gap state at 2.4~eV below the conduction band reported there is not unambigously 
visible in our data. Such a level is only present in LAO/TiO$_2$:STO and -- astonishingly -- 
in the bare STO with mixed termination.

The discussion of the features above the STO bandgap appears to be more problematic, since 
already the overall shapes of the spectra are hardly comparable (see again 
figure~\ref{LAO_STO_comp_spectra_super}). 
Furthermore, the SPV transient behavior of STO under illumination with super-bandgap light is 
characterized by multiple carrier transfer processes, which take place at long time scales of 
minutes or even hours \cite{bey13}. As a consequence, spectral features might interfere with 
transient features and the SPV spectra can hardly be interpreted as a true energetic map 
of electronic states. Certainly, higher optical transitions in the STO play a 
decisive role in this spectral range, while the following point suggests that
in-gap states communicating with the LAO bands are of minor importance: 
If the LAO single crystal is taken as reference (figure~\ref{LAO_spectra}), 
two levels at 4.9~eV and 5.2~eV should 
be observable exclusively in the LAO-covered samples. 
However, at 4.9~eV no point of slope change was observed, 
while for 5.1/5.2~eV indeed a level is found -- but also for the 
bare STO crystals, which cannot be exhaustively explained at the current state of knowledge.  

\section{Summary and outlook}
\label{Summary}

Spectral and time-resolved measurements of the surface photovoltage were employed to 
compare two 5-u.c.-thick LaAlO$_3$ films 
on either TiO$_2$- or SrO-terminated SrTiO$_3$ as well as 
two bare SrTiO$_3$ single crystals with the respective surface terminations with regard to 
their distribution of electronic interface states. The detailed analysis of the sub-bandgap 
SPV spectra showed a quite individual defect state map for each of the four samples. However, there 
were very clear similarities in (i) the sign and the shape of the sub-bandgap spectra as well as 
(ii) the shape of the light-off SPV transients at 600~nm excitation wavelength between the 
two TiO$_2$-terminated samples and the two SrO-terminated ones, respectively. 
This result is in accordance 
with a number of recent publications, in which the creation of a conducting 
interface did not necessarily require a crystalline LaAlO$_3$ film 
\cite{che12,lee12,del12,chr13}, but seemed to be crucially 
dependent on the physics of the SrTiO$_3$ substrate. Still, this remains a 
key point of the debate, since at the same time there is strong evidence that 
the LAO stoichiometry plays a decisive role for the creation of the conducting 
electron liquid \cite{war13}.

In order to improve the microscopic picture of the charge dynamics 
in oxide heterostructures including a less speculative association 
of the SPV spectral features to 
bulk- or interface related electronic trap states,
the collection of further data, e.g. on 
LAO/STO samples with gradually changed LAO thickness and 
with gradually changed oxygen 
vacancy concentration is indispensable. Such 
investigations should becarried out in an ultrahigh-vacuum 
environment to eliminate extrinsic SPV contributions, which 
typically complicate the shape of the 
SPV transients \cite{fou09b}.

\begin{acknowledgments}
This work was financially supported by the German Research 
Foundation (BE 3804/2-1 and EN 434/31-1). We thank Claudia Cancellieri, Stefano Gariglio, and 
Jean-Marc Triscone for providing us with samples and for invaluable discussions.
\end{acknowledgments}


%

\end{document}